\documentclass[sigconf]{acmart}
\renewcommand\footnotetextcopyrightpermission[1]{} 
\usepackage{balance}
\usepackage{color}

\newif\ifworkinprogress
\workinprogresstrue

\ifworkinprogress
  \newcommand{\mq}[1]{\textcolor{blue}{\textbf{[Massimo] #1}}}
  \newcommand{\es}[1]{\textcolor{magenta}{\textbf{[Erik] #1}}}
  \newcommand{\mr}[1]{\textcolor{cyan}{\textbf{[Marta] #1}}}
  \newcommand{\ty}[1]{\textcolor{red}{\textbf{[Tao] #1}}} 		   \newcommand{\pv}[1]{\textcolor{green}{\textbf{[Puya] #1}}}
\else
  \newcommand{\mq}[1]{}
  \newcommand{\es}[1]{}
  \newcommand{\mr}[1]{}
  \newcommand{\ty}[1]{}
  \newcommand{\pv}[1]{}
\fi

\usepackage{csquotes}

\usepackage{epsfig}
\usepackage{color}
\newcommand{\hl}[1]{\colorbox{yellow}{#1}}
\usepackage{balance}  
\usepackage{algorithmic}
\usepackage{algorithm}
\usepackage{multirow}
\usepackage{graphicx}
\usepackage{amsmath}
\usepackage{amssymb}
\usepackage{amsfonts}
\usepackage{xspace}
\usepackage{url}
\usepackage{./style/tweaklist}
\usepackage{hyperref}
\usepackage{booktabs}
\usepackage{bbold}
\newcommand{\normaltilde}{\raise.17ex\hbox{$\scriptstyle\sim$}}

\newcommand{\IGNORE}[1]{}

\newcommand{\RNNt}{\textit {Recurrent Neural Network}}
\newcommand{\RNN}{\textit {RNN}}

\newcommand{\KNN}{\textit {KNN}}

\newcommand{\WTOVEC}{\textit {W2V}}

\newcommand{\POP}{\textit {POP}}

\newcommand{\commentedtext}[1]{}

\usepackage{booktabs} 

\graphicspath{{./figures/}}

\begin{document}

\title{Modeling Musical Taste Evolution \\with Recurrent Neural Networks}


\author{Massimo Quadrana}
\affiliation{%
  \institution{Pandora Media Inc.}
  \streetaddress{}
  \postcode{}
}
\email{mquadrana@pandora.com}

\author{Marta Reznakova}
\affiliation{%
  \institution{Pandora Media Inc.}
  \streetaddress{}
  \postcode{}
}
\email{mreznakova@pandora.com}

\author{Tao Ye }
\affiliation{%
  \institution{Pandora Media Inc.}
  \streetaddress{}
  \postcode{}
}
\email{tye@pandora.com}

\author{Erik Schmidt}
\affiliation{%
  \institution{Pandora Media Inc.}
  \streetaddress{}
  \postcode{}
}
\email{eschmidt@pandora.com}

\author{Hossein Vahabi}
\affiliation{%
  \institution{Pandora Media Inc.}
  \streetaddress{}
  \postcode{}
}
\email{puya.vahabi@gmail.com}


\begin{abstract}
Finding the music of the moment can often be a challenging problem, even for well-versed music listeners.  Musical tastes are constantly in flux, and the problem of developing computational models for musical taste dynamics presents a rich and nebulous problem space.  A variety of factors all play some role in determining preferences (e.g., popularity, musicological, social, geographical, generational), and these factors vary across different listeners and contexts. In this paper, we leverage a massive dataset on internet radio station creation from a large music streaming company in order to develop 
computational models of listener taste evolution.  We delve deep into the complexities of this domain, identifying some of the unique challenges that it presents, and develop a model utilizing recurrent neural networks. We apply our model to the problem of next station prediction and show that it not only outperforms several baselines, but excels at long tail music personalization, particularly by learning the long-term dependency structure of listener music preference evolution.
\end{abstract}

%
%


%
%

\maketitle

\section{Introduction}
\label{sec:introduction}
%
Musical preference dynamics present a rich and nebulous problem space for the development of models capable of predicting the temporal structure of user behavior.  Music listening is an extremely long-tailed distribution, where a massive proportion of user behavior is dominated by an industry that's constantly delivering new pop stars in addition to the ebb and flow of popularity among mega-artists with careers spanning multiple decades. 

However, outside of this ``short head'' behavior, there exists a majority of listeners who have interest spanning over hundreds of thousands of artists and millions of tracks.  Their music discovery is influenced by a multitude of factors such as social, geographical, generational, or musicological.  They may be going through a maturation of preference in a discovery of 1950's bop jazz or perhaps developing a guilty pleasure in 1990's pop music.

Despite the richness and complexity of this domain, music preference has received very little attention in the recommendation literature to date.  At the time of writing, the authors are only aware of one other paper that has addressed the topic \cite{moore13taste}.  

In this paper, we set out to provide a deep dive into this topic and to develop models which are capable of encoding the evolution of listener music preferences.  Our ultimate goal is to provide extremely powerful recommendations, but this must be achieved through the ability to predict where their preference is headed next.

To construct our dataset of musical preferences, we leverage internet radio data from Pandora, a US-based music streaming service with over 80 million monthly active listeners.  We compile a dataset of personalized radio station creation, sampling from a dataset of over 11 billion stations that have been created to date.  These stations can be created (or ``seeded'') from an artist, track, composer or genre, and become personalized to the listeners tastes as they provide feedback. 

Listeners add stations to their profile strictly in a sequential manner over time, in synchrony with the evolution of their tastes. Modeling station creation over time allows us to look at the dependency structure over these potentially large changes in listening behavior. 



We focus directly predicting a user's next search-created stations, based on past behavior. 
We propose a variety of model options, focusing extensively on the use of recurrent neural networks (\RNN s).  \RNN s have been successfully used throughout academic and commercial applications \cite{sordoni2015hierarchical,lipton2015critical,hidasi16session} and are powerful in sequential applications as they allow us to model complex temporal dependency structure. We compare this to a set of baseline models based on popularity and industry standard similarity, and show where \RNN~models are most powerful.

Given these trained models, we do a deep dive in order to understand the importance of past data in predicting a listener's future music preferences.  We segment our dataset into various cohorts by both historical and future station creation counts as well as station popularity. This slice and dice method helps us identify model strength and weaknesses in the prediction task. Finally, we experiment with combining our ultimate \RNN~model in a cascaded learn-to-rank model which demonstrates that the model delivers strong information gain over the baselines.


The remainder of this paper is organized as follows: In Section \ref{sec:related} we give an overview of related work. In Section \ref{sec:model} we present our model and other techniques that can be used in this context. In Section \ref{sec:experiments} we describe our experimental framework and the dataset. We share key experiment results in Section \ref{sec:results}. We conclude in Section \ref{sec:conclusion} with a discussion of the results and future work.

\section{Related work}
\label{sec:related}
%
Modeling user preference dynamics over time has been the subject of investigation across multiple recommendation domains \cite{koren09cf,moore13taste}.
In recommender systems, modeling the evolution of user tastes over time is crucially important in order to to react to drift in the user's preferences and to adapt recommendations accordingly. 

\subsection{Time-Aware Recommender Systems}
Time-aware recommender systems (TARS) \cite{campos14time} exploit time as an additional contextual dimension that is added to traditional recommender systems. In these systems,   ratings timestamps are exploited to identify periodicity in user habits. In general, the goal is to improve the accuracy of collaborative filtering models like matrix factorization by adding temporal dynamics to otherwise static profiles \cite{campos11towards,koren09cf}. A downside to this family of models is that they are dependent on the exact timestamps at which user feedback is collected, which in turn may not correspond to the actual consumption time. Moreover, users who have a similar evolution in their tastes over different temporal scales will be treated differently by these models.

\subsection{Sequence-Aware Recommender Systems} \label{sec:sars}
Sequence-aware recommender systems (SARS) provide the necessary flexibility by relaxing the aforementioned temporal constrains and focusing exclusively on the sequential order of the historical user activity. 

SARS models have been shown to be beneficial in scenarios such as e-commerce \cite{jannach15adaptation}, news \cite{garcin13news}, and video recommendation \cite{hidasi16feature}. They are particularily powerful in doimains in which the sequence of user's recent interactions has a greater predictive power over their historical profile in terms predicting which items they will interact with in the future.

Moreover, sequential user behavior becomes crucial in session-based scenarios in which no user profile can be constructed because of the absence of historical feedback.  In these cases, it is vital to draw as much information as possible from current session \cite{hidasi16session}.
%

Sequence-aware recommendation is typically tackled by extracting sequential patterns from logs of historical user activity that are later matched to the recent or historical user actions to generate recommendations \cite{hariri12context,mobasher02sequential}. Alternative solutions used Markov models to explicitly model the transition probabilities between users' states \cite{shani05mdp}.

However, both models suffer from the inherent data sparsity in the interaction space. With massively huge catalogs of items the state space quickly becomes unmanageable when trying to include all possible sequences of potential user selections over all items, and the frequency of each user selection can potentially be really small to invalidate the frequent pattern extraction procedure. The impact of data sparsity can be alleviated by combining Markov models with latent factor models \cite{rendle10FPMC}, but its applicability is still limited by the dimension of the state space.

\subsection{Deep Learning in Recommender Systems}
The recent explosion of interest in deep learning models has opened new frontiers for sequence modeling when large amounts of data are available. Recurrent neural networks (\RNN s) have achieved dramatic improvements over item-to-item models in session-based recommendation, even when the content of items was explicitly modeled into the sequence of interactions \cite{hidasi16session,hidasi16feature}. Word embeddings models \cite{mikolov2013distributed} have been used to generate sequence-based item representations that can be later used with classical item-similarity models to provide sequential recommendations \cite{grbovic15prod2vec,vasile16meta}.

\RNN s are the deep models of choice when dealing with sequential data \cite{lipton2015critical}. \RNN s have been used in image and video captioning, time series prediction, natural language processing, query recommendation \cite{sordoni2015hierarchical} conversational models, text and music generation, and much more. Long short-term memory (LSTM) \cite{hochreiter1997long} networks are a type of \RNN s that have been shown to work particularly well, it includes additional gates that regulate when and how much to take the input into account and when to reset the hidden state. This helps with the vanishing gradient problem that often plagues the standard \RNN~ models. Slightly simplified version of LSTM -- that still maintains all their properties -- are gated recurrent units (GRUs) \cite{cho2014properties} which we use in this work.

Furthermore, \RNN s are not affected by the data challenges of the SARS models discused in \ref{sec:sars} and can be easily scaled to datasets of the size of used by music streaming platform. 

\subsection{Music Domain Applications}
In the music domain, sequential features extracted from historical listening records can be used to generate coherent continuations to the current listening session or playlists \cite{bonnin14playlists,jannach15playlist}.

To our knowledge, only \cite{moore13taste} considered a sequence based approach to modeling the evolution of user tastes through time in music. User and song transitions are embedded into a common latent first-order Markov space that can browsed with Gaussian random walks. However, this approach is affected by the same limitation in the size of the state space of Markov models, which in turn severely constraints the number of users and songs that can be effectively represented. 

This work is the first one that uses \RNN s for modeling user preference dynamics in music to our knowledge.

\section{Model}
\label{sec:model}
%

In this section we describe our model (and other methods) to capture the musical taste evolution of the listeners and in particular to solve the problem of listener's next search-created stations. 

\subsection{Recurrent Neural Network}
\begin{figure}[t]
\centering
\includegraphics[width=1\linewidth]{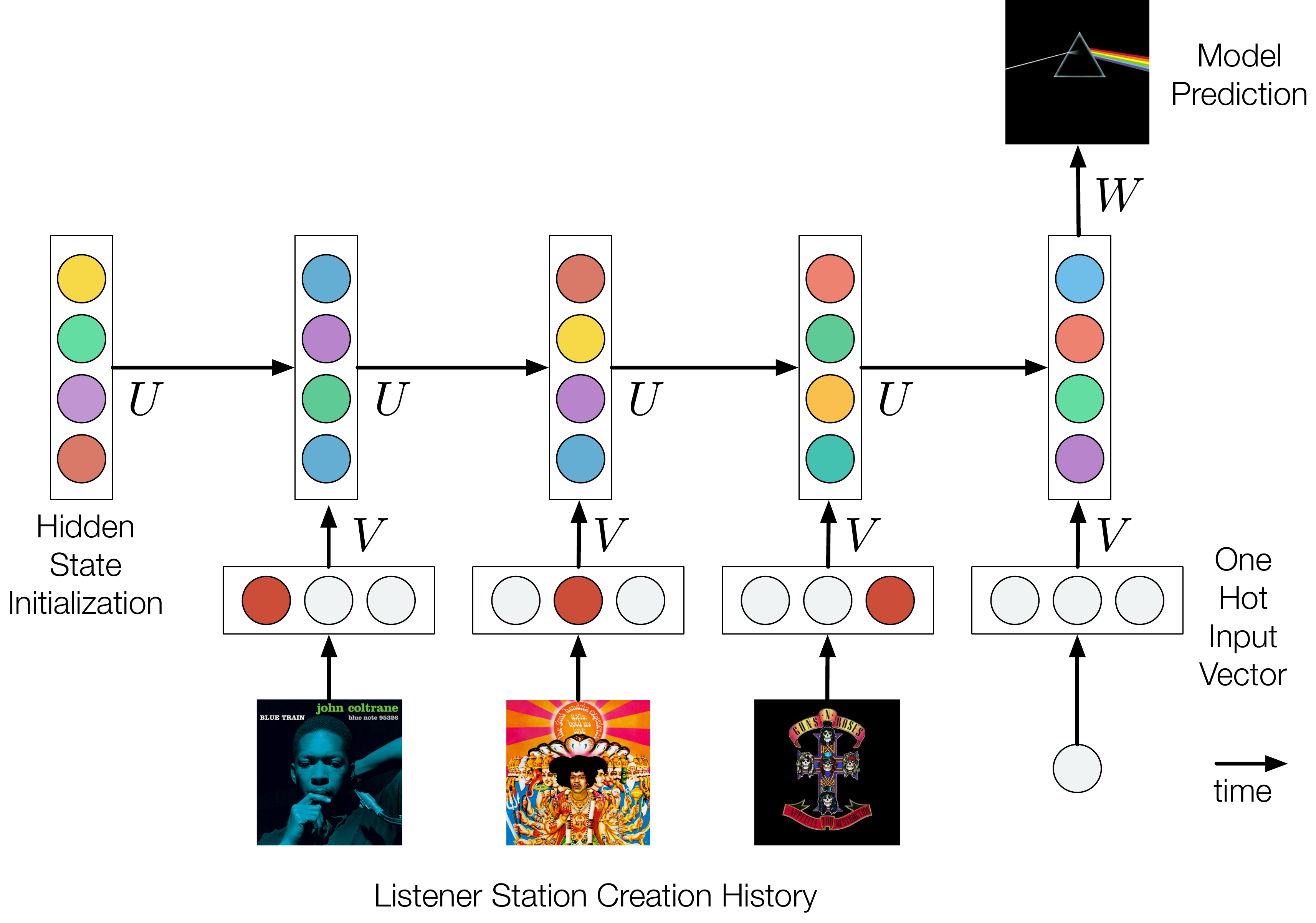}
\caption{\RNN~ model at a high level.  The red dots in the input layer represent the one-hot encoding of the input and the colors in the hidden layer represent the model activations.}
\label{fig:rnn}
\end{figure}

The \RNNt\ (\RNN) computes for each station $s_t$ a dense vector $h_t$, the recurrent state, that combines the previous $s_t$ with the previous state $h_{t-1}$ according to the formula,
\begin{equation}
h_t = f(s_t, h_{t-1}), h_0 = 0,
\end{equation}
where $h_t \in \mathbb{R}^{d_t}$, $d_t$ is the number of dimensions of the recurrent state, and $f$ is a non-linear transformation.
In this work we adopted the Gated Recurrent Unit (GRU) model for the modeling of click-stream data described in \cite{hidasi16feature}. Each station $s_n$ is represented as one-hot vector, i.e., a vector of length equal to the number of stations $S$ with a 1 corresponding to the index of the current station, and 0 otherwise. The parametrization $f$ of GRU is given by:
\begin{eqnarray}
r_t = \sigma\left(U_r s_t + W_r h_{t-1}\right), &\text{(\textit{reset gate})}\nonumber\\
z_t = \sigma\left(U_z s_t + W_z h_{t-1}\right),&\text{(\textit{update gate})}\\
\tilde{h}_t = \mathrm{tanh}\left({U_h z_t + W_h(r_t \cdot h_{t-1})}\right),&\text{(\textit{candidate update})}\nonumber\\
h_t = (1-z_t)h_{t-1} + z_t\tilde{h}_t,&\text{(\textit{final update})}\nonumber
\end{eqnarray}
where $\sigma$ is the logistic sigmoid, $\cdot$ is the element-wise scalar product between vectors, $U_r,U_z,U_h \in \mathbb{R}^{d_h \times S}$ and $W_r,W_z,W_h \in \mathbb{R}^{d_h \times d_h}$. The GRU predicts the next station in the sequence according the following,
\begin{equation}
\hat{s}_{t+1} = g(h_{t-1}),
\end{equation}
where $g$ is another non-linear transformation that projects the recurrent representation of the GRU onto the space of the stations. The formulation of $g$ depends on the loss function used to train the GRU. In accordance to \cite{hidasi16feature}, we initially experimented with three different loss functions: cross-entropy, Bayesian Pairwise Ranking (BPR) and TOP-1 loss. With cross-entropy $\hat{s}_{t+1} = \mathrm{softmax}(W_y h_{t-1})$, whereas with BPR and TOP-1 $\hat{s}_{t+1} = \mathrm{tanh}(W_y h_{t-1})$. We settled on BPR loss function since it yields the best results. A graphical description of this model is provided in Figure \ref{fig:rnn}.

The GRU is trained to predict the next station given the complete past history of the user that is encoded into its recurrent state $h_t$. The training is performed end-to-end such that sequential relationships between stations are learned directly from data without the need of ad-hoc feature extraction.

In our experiments, we used a single layer GRU network trained with parallel mini-batches and negative output sampling. These features are of crucial importance for training the model over the large dataset used for our experiments.
 
\subsection{Similarity-based baseline models}
In addition to the \RNN, we have considered several similarity-based methods for next-station prediction. These methods typically assign a score to each candidate next-station that is proportional to its similarity with the previous station in the sequence. 
In this work, we have defined item similarity wrt.\ co-occurrence of stations in user profiles. Specifically, we have considered the overall item popularity, co-occurrence \cite{hidasi16session} and Word2Vec \cite{grbovic15prod2vec} based similarities. Despite their apparent simplicity, these models turn out to be very strong baselines in this domain, as we will show in the experimental evaluation.

\subsubsection{Item Popularity (\POP)}
Given the extremely long tailed distribution of music, it's extremely important to consider a popularity baseline for any recommendation approach.  For this baseline, we simply take the most popular station seeds (i.e., artist, genre, composer, track) across the service and utilize the top K as the recommendations.

\subsubsection{Co-occurrence kNN.(\KNN)}
A simple way of assessing the similarity between items is to consider how frequently they co-occur.  More concretely, we compute how many times two different items occur in the same sequence of user interactions. This rather simple approach, that leads to recommendations of the type ``others who added this station also added these other ones,'' has proven its effectiveness in a variety of domains, most notably in e-commerce \cite{linden2003amazon}.

In this work, we have to the co-occurrence item-item similarity metric used in \cite{hidasi16session} for next-click recommendation in click-stream data. The similarity between two items $i$ and $j$ is computed as,
\begin{equation}\label{eq:knn_sim}
\textrm{sim}(i,j) = \frac{|S(i) \cap S(j)|}{\sqrt{|S(i)| + \lambda} \cdot \sqrt{|S(j)| + \lambda}},
\end{equation}
where $S(z)$ returns the sessions (sequences) in the training database where the item $z$ occurs at least once, and $\lambda$ is a damping factor to avoid coincidental high similarities of rarely visited items.
At recommendation time $T$, all items are ranked according to their similarity w.r.t.\ the last item in the sequence $i$, and then the top-$k$ similar items to the last one in the sequence are recommended to the user. 

The damping factor is utilized in order to mitigate the impact of data sparsity on the similarity metric, which is otherwise strongly constrained by the actual co-occurrences that are observed in the data. Despite these challenges, this approach showed very strong performance in our experiments.

\subsubsection{word2Vec kNN.(\WTOVEC)}
In addition to item co-occurrence, more complex interactions between items can be modeled from the sequences of user interactions. In particular, we adapt \textit{prod2vec} \cite{grbovic15prod2vec} to the station recommendation case. The \textit{prod2vec} model involves learning vector representations of products from sequential actions of users.
In our case, we learn a vector-wise representation of stations from the sequence of station created by users. In particular, we consider the sequence as ``sentence'' and single station as ``words.'' Further details on this method can be found in \cite{grbovic15prod2vec}. 


\section{Experiments}
\label{sec:experiments}
%
\begin{figure}[thb]
\begin{center}
\includegraphics[width=.9\linewidth]{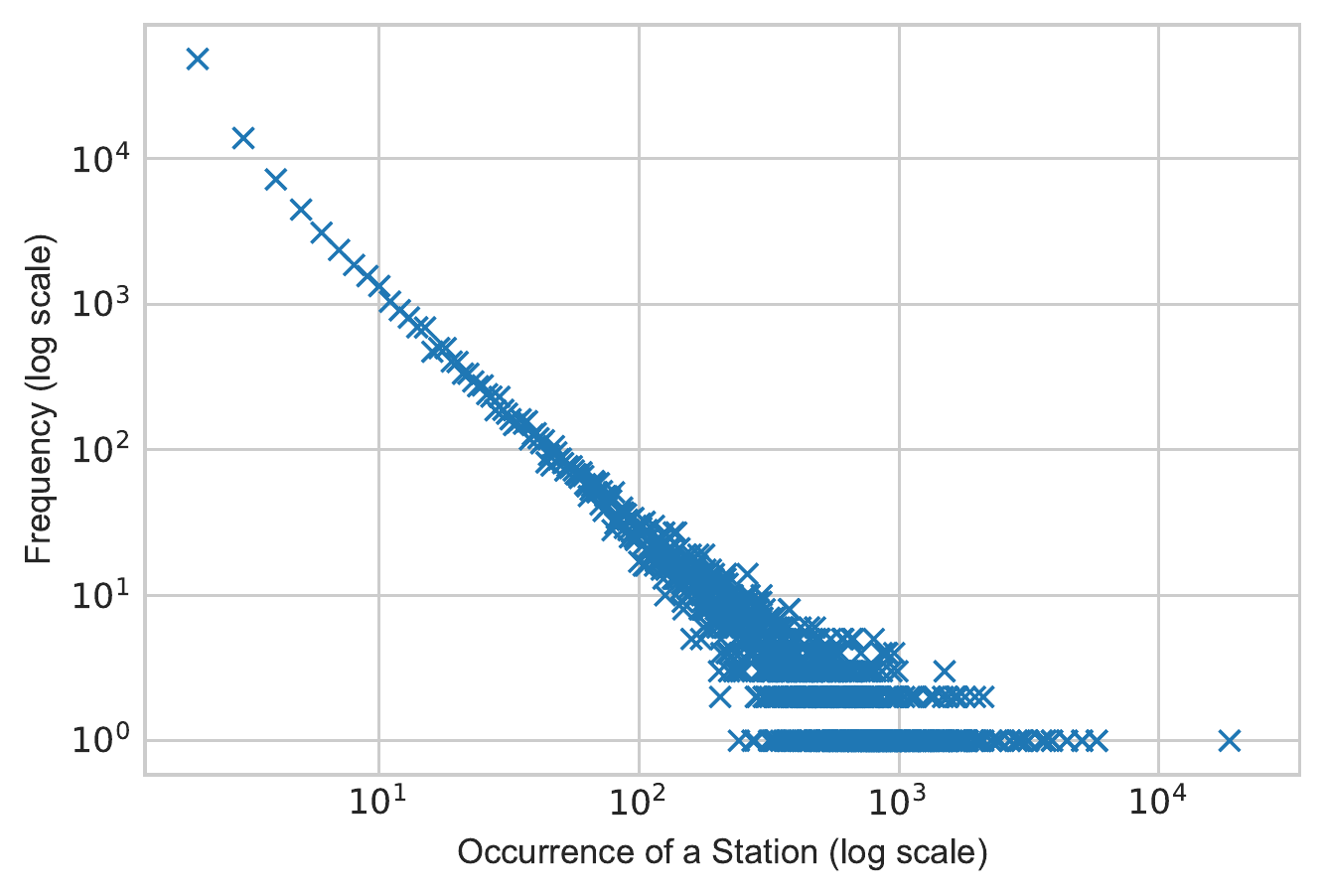}
\caption{The distribution of stations by popularity in log log. The occurrence of user created stations represents popularity of a station, in x-axis. The number of such stations is shown in the y-axis. Very popular stations such as Today's Hits Radio has a high occurrence in creation, but there exists only a few of these stations. }
%
%
\label{fig:dist_stations}
\end{center}
\end{figure}

In the following experiments we will investigate employing a variety of models for predicting listener preference dynamics.  After an explanation of the dataset and metrics, we will evaluate our model in the task of predicting the next stations that a listener will create.  


\subsection{Dataset}
\label{subsec:dataset}

For our experiments, we have assembled a data collection containing the stations added by a random sample Pandora listeners over a 1-year period. We consider the stations together with the time-stamp in which they were created. To reduce the impact of selection bias \cite{Schnabel16bias}, i.e. the existing recommendation systems used by the platform to suggest new stations to listeners, we considered only stations added through search-box. While it is still possible that listener choices are be influenced by recommendations received in other parts of the application (e.g., bias present in the process of track selection on an existing station exposing a listener to a new artist), searches through the search-box are clear expressions of the intent of the listener of looking for a new station. These are less-biased signals that we believe are good indicator of listener interests and hence adequate for modeling the actual interests of the listener and their evolution through time.


\begin{table}[h]
\centering
\begin{tabular}{ l | r } 
 \toprule
 Number of Listeners & $330,170$ \\ 
 Number of Stations & $3,079,399$ \\
 Number of Listeners (Training) & $324,750$  \\
 Number of Stations (Training) & $2,912,564$  \\
 Number of Listeners (Test) & $64,750$  \\
 Number of Stations (Test) & $151,015$  \\ 
 \bottomrule
\end{tabular}
\caption{Statistics of the Dataset.}
\label{tab:dataset}
\end{table}

Details about our dataset can be found in \autoref{tab:dataset}. We collected sequences of stations added by \normaltilde$300$k listeners over a 1-year period. \autoref{fig:dist_stations} depicts the distribution of station creation. As previously discussed in \autoref{sec:introduction}, music popularity follows a power-law distribution where a very small number of stations receive an overwhelming share of total creation. 

To create training and testing datasets, we partitioned our dataset temporally, considering the first 11 months for training and the last month for testing. 

As our ultimate focus is to predict how a listener's taste will evolve, we focus exclusively on listeners who have created a minimum of three stations on the service.  Listeners who have created only one or two stations are filtered from our dataset.  Furthermore, to remove spurious stations from our training dataset we also removed any station which was listened for less than one hour. This leads to a slight overall reduction of the number of stations both through this filtering as well as our restriction to include only stations created via listener searches. The final training/testing dataset sizes are reported in \autoref{tab:dataset}.
 

\subsection{Experimental Setup}
Given a set of stations created by the listener in the training set $u_{train}$ = ($s_1$, $s_2$, $s_3$, \ldots, $s_t$) and a set of stations created by the same listener in the test set $u_{test}$ = ($s_{t+1}$, \ldots, $s_n$), in order to evaluate our algorithms we aim to predict the test stations $u_{test}$. As evaluation metrics we use standard information retrieval measures. For each listener we compute:
\begin{itemize}
\item \emph{MRR}: Mean reciprocal rank is the inverse of the position of the first correct station in the ranking produced by the algorithm.
\item \emph{Recall@K}: The fraction of stations correctly predicted in $K$ suggestions divided by overall number of stations created by the listener in the test set. 
\end{itemize}

\subsubsection{RNN Parameter Tuning}
Our models are implemented based python package Theano\footnote{\url{http://deeplearning.net/software/theano/}} 0.9.0, and experiments are performed on NVIDIA Tesla M40 GPUs. As mentioned in Section \ref{sec:model}, we used Bayesian Pairwise Ranking as the loss function. For each size of the hidden units we tuned a set of parameters using random search. The tuning was performed on a validation set that is of the same nature as the test set, i.e. contains a set of listeners with an ordered sequences of stations created. We have tuned the networks for Recall@$10$. We found the optimal learning rate to be $0.05$ for all the unit sizes. We set the  momentum to $0.1$, and dropout to $0.1$. 

In order to understand the impact of the hidden unit size on effectiveness, we trained initially the network with $3$ different unit size: $200$, $500$, $1000$. 
\begin{table}[thb]
\centering
\begin{tabular}{l|l|l|l|l}
\toprule
& @$5$ & @$10$ & @$20$ & @$50$\\
\midrule
$200$ & $0.0264$ & $0.04406$ & $0.07077$ & $0.1273$\\
$500$ & $0.02753$ & $0.04599$ & $0.07476$ & $0.1367$\\
$1000$ & $0.02819$ & $0.04831$ & $0.07812$ & $0.1378$\\
\bottomrule
\end{tabular}
\caption{Recall@K for different hidden unit sizes.}
\label{tab:rnn_units_comp}
\end{table}
Table \ref{tab:rnn_units_comp} reports Recall@$K$ metrics for different sizes of hidden units. Recall increases with more units for all the $K$. However, units larger than $1000$ led to computational cost limitations without significant performance gains. We choose $1000$ units for the rest of the experiments.

\section{Results}
\label{sec:results}
\begin{figure}[tb]
\centering
\includegraphics[width=.9\linewidth]{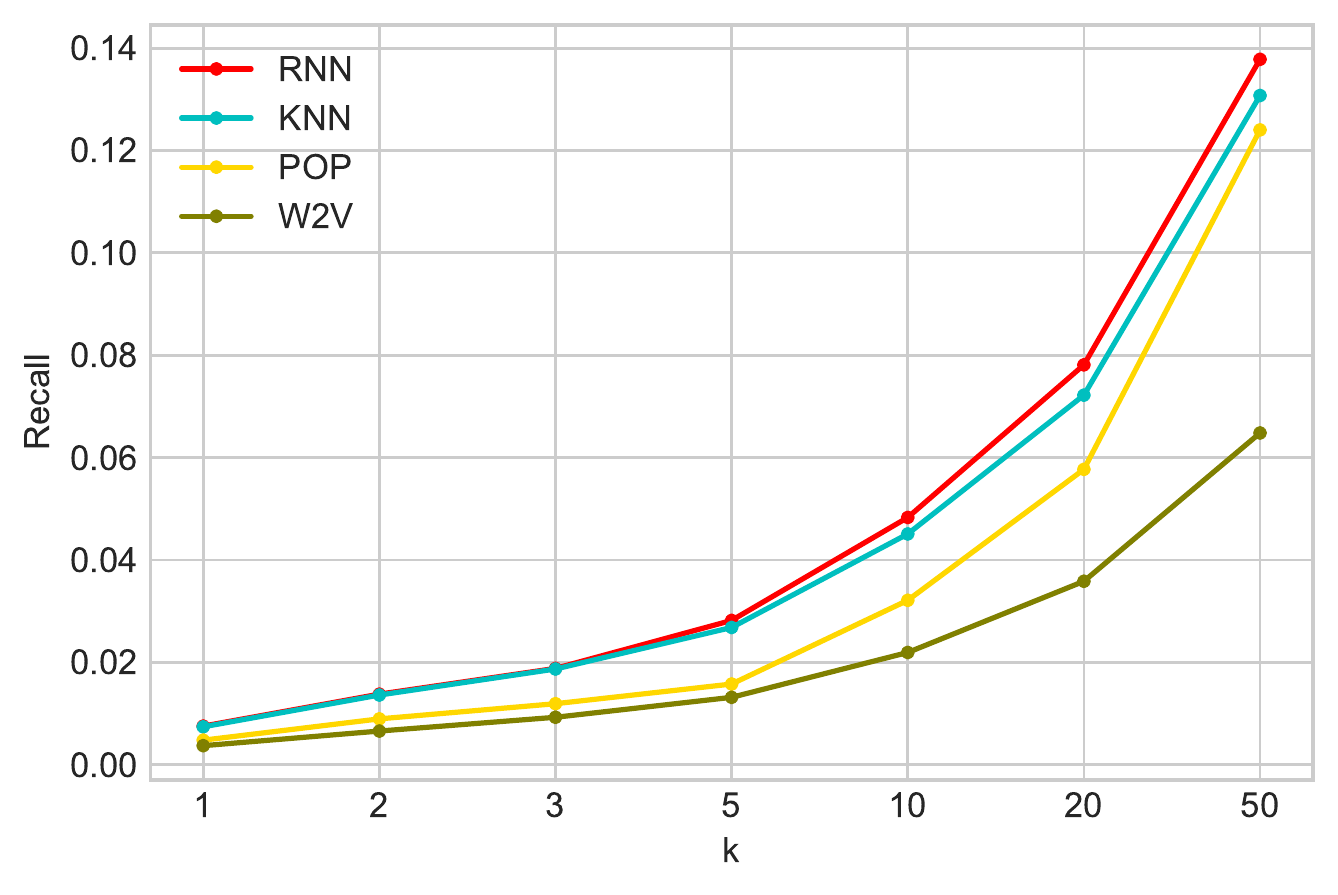}
\caption{Recall@K \RNN~ and baseline algorithms.}
\label{fig:recall_overall}
\end{figure}

\begin{figure}[tb]
\centering
\includegraphics[width=.9\linewidth]{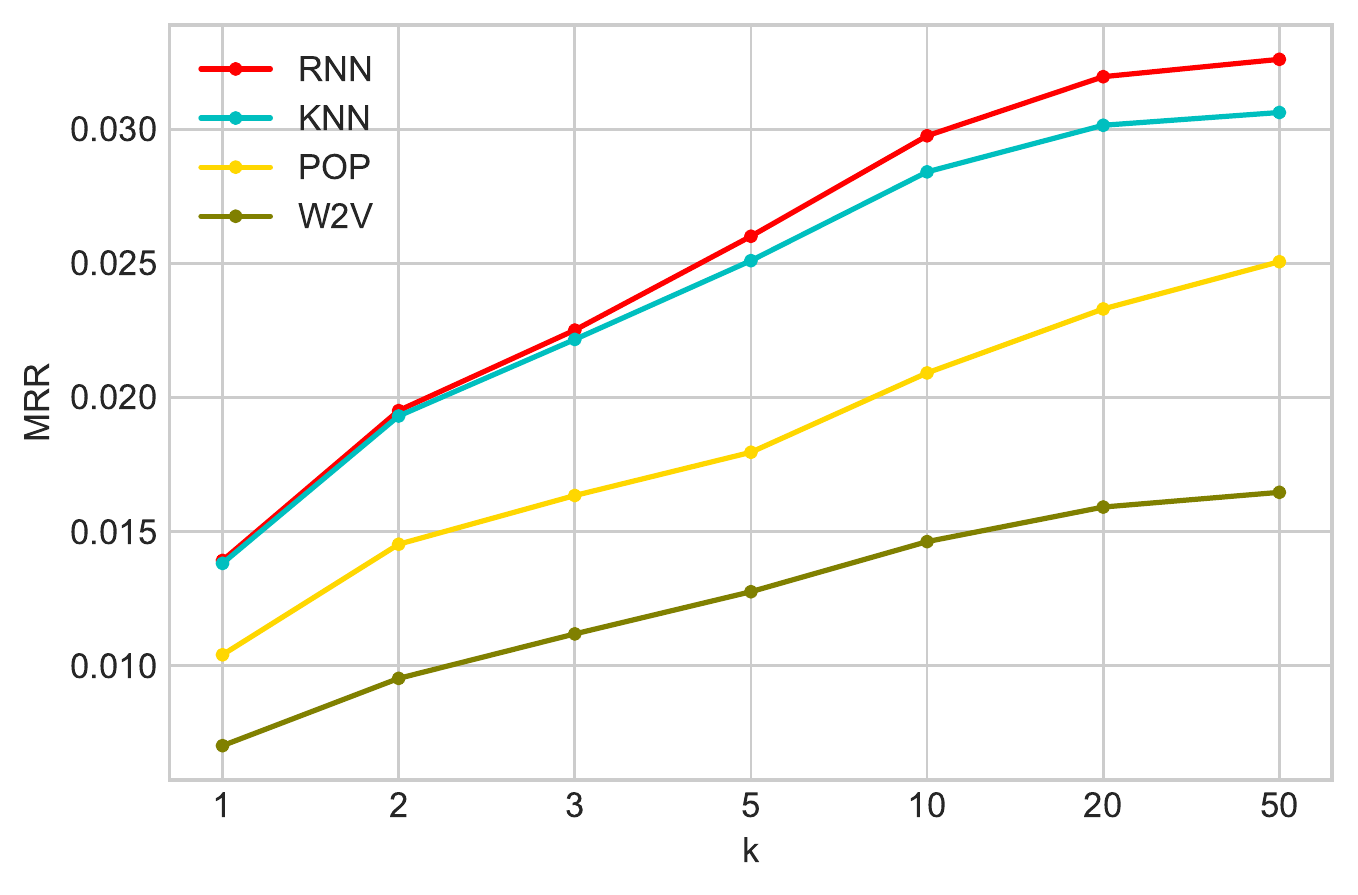}
\caption{MRR@K \RNN~ and baseline algorithms.}
\label{fig:mrr_overall}
\end{figure}

We report general results of next search-created stations prediction, in Recall in Figure \ref{fig:recall_overall}, and MRR in Figure \ref{fig:mrr_overall}, for different prediction cutoff $K$ ranging from $1$ to $50$. In all values of $K$, \KNN~ is the best performing baseline model, outperforming popularity (\POP), and word2vec (\WTOVEC) by a large margin.

\RNN~ outperforms the best baseline model \KNN~ when $K > 10$, with +$8.20\%$ in Recall@$20$, and $+5.41\%$ in Recall@$50$. For small $K$ ($\le 10$), there is no big advantage in recall. However, in MRR metric, even at small $K = 5$, and $K = 5$, \RNN~ ranks better than \KNN~ by $1.57\%$ and $3.62\%$, respectively.  In Figure \ref{fig:mrr_overall}, we can see that with increasing $K$ value, \RNN's margin of improvement over \KNN~ also increases consistently in MRR. Detailed result comparisons of \RNN~ and \KNN~ are reported in  \autoref{tab:rec_full}.

While we expected the \RNN~ model to predict next stations better because it specifically modeled sequential behavior of listener's taste, we want to dive deep in listener analysis to substantiate this claim. Typically this is done through slicing and dicing based on listener characteristics.


\begin{table}[t]
\centering
\begin{tabular}{l|l|l|l|l|l}
\toprule
\multicolumn{6}{c}{Recall} \\
\midrule
& @$1$ & @$5$ & @$10$ & @$20$ & @$50$\\
\midrule
\KNN~ & $0.007401$ & $0.02681$ & $0.04509$ & $0.0722$ & $0.1307$\\
\RNN~ & $0.007538$ & $0.02819$ & $0.04831$ & $0.07812$ & $0.1378$\\
$\Delta$ \% & $+1.9\%$ & $+5.1\%$ & $+7.1\%$ & $+8.2\%$ & $+5.4\%$\\
\midrule
\multicolumn{6}{c}{MRR} \\
\midrule
\KNN~ & $0.01382$ & $0.02511$ & $0.02842$ & $0.03016$ & $0.03063$\\
\RNN~ & $0.01393$ & $0.02602$ & $0.02976$ & $0.03197$ & $0.03261$\\
$\Delta$ \% & $+0.8\%$ & $+3.6\%$ & $+4.7\%$ & $+6.0\%$ & $+6.5\%$\\
\bottomrule
\end{tabular}
\caption{Overall Recall@K and MRR@K for \RNN~ and the best baseline \KNN.}
\label{tab:rec_full}
\end{table}

\subsection{Listener Segmentation Study}
\begin{figure}[thb]
\includegraphics[width=.9\linewidth]{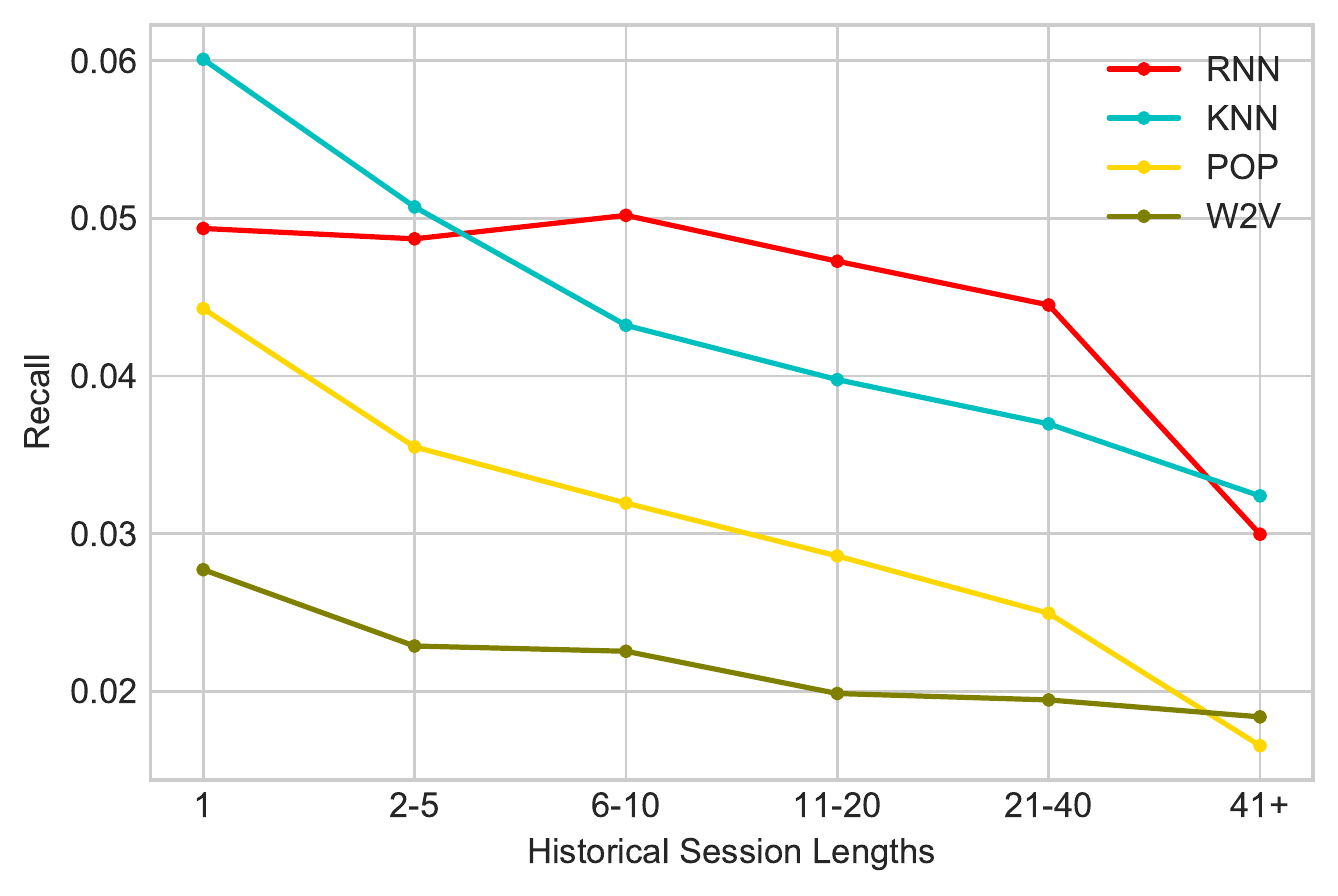}
\caption{Recall@10 \RNN~ and baseline algorithms, segmented by historical session length.}
\label{fig:recall_by_historical}
\end{figure}

\begin{figure}[thb]
\includegraphics[width=.9\linewidth]{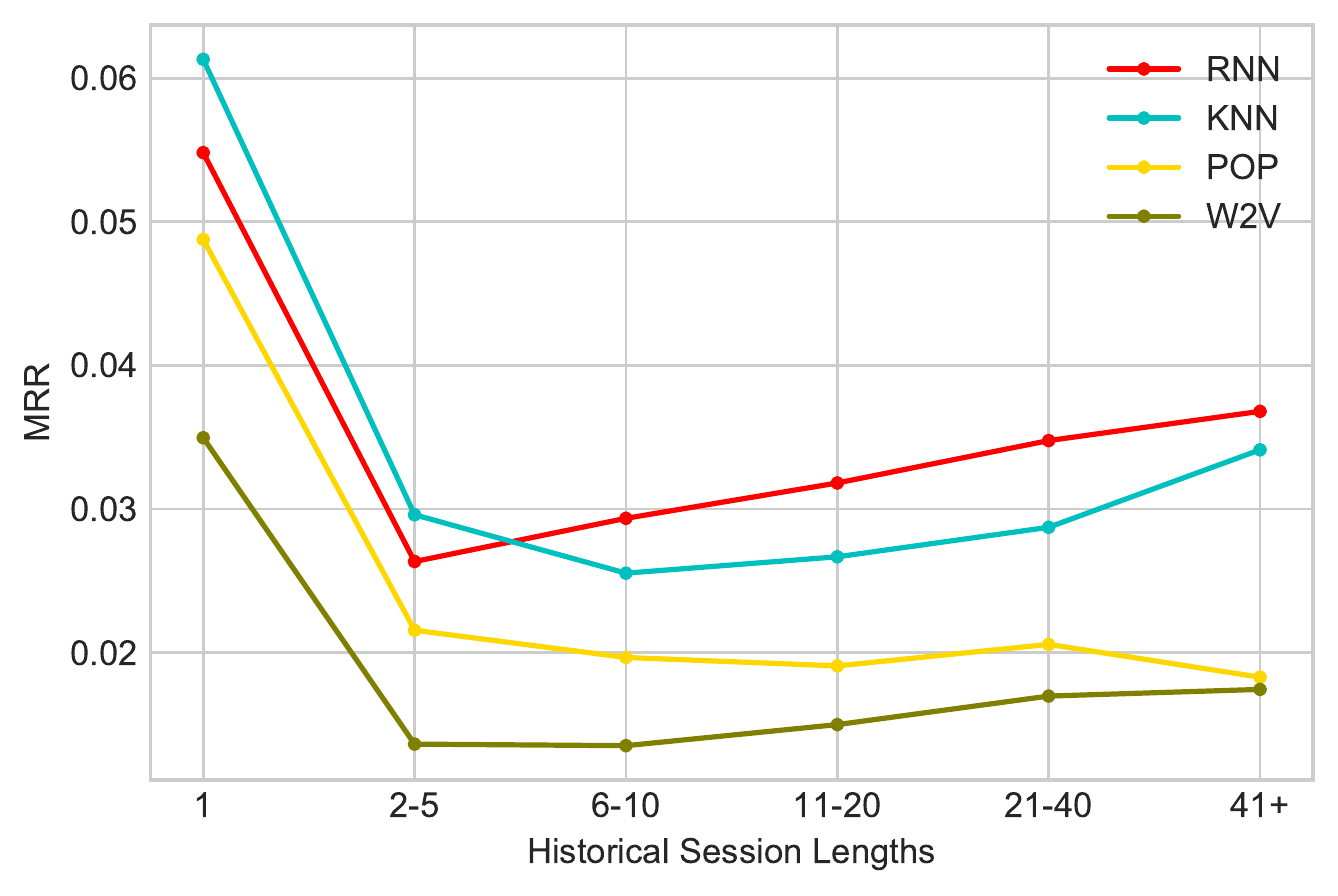}
\caption{MRR@10 \RNN~ and baseline algorithms, segmented by historical session length.}
\label{fig:mrr_by_historial}
\end{figure}

We perform several listener segment studies. In this study we fix $K=10$ for both Recall and MRR metrics. First, we segment listeners based on their historical session length (number of stations created by a listener in the training set), represented by the number of previously created stations before the test period. We further bin listeners into 1, 2-5, 6-10, 11-20, 21-40, and 41+ previous stations. The fewer stations a listener has, the less listening history data we have. Figure \ref{fig:recall_by_historical} and Figure \ref{fig:mrr_by_historial} show that both \RNN~ and \KNN~ outperforms \POP~ and \WTOVEC~ in all listener segments in Recall@10 and MRR@10. For listeners with very short history, i.e., $\le 5$ stations, \KNN~ is the best algorithm. However, for listeners with more than 5 stations, i.e. when sequential patterns might be emerging, \RNN~ algorithm overtakes \KNN~ by 11.9\% in Recall@10. We attribute this to neural network's ability to retain memory. This suggests that our 1 layer network structure is very efficient at learning preference changes when given enough history. Yet it is not wise to use it as a standalone model as \KNN~ clearly claims an advantage with listeners with a very short history. More in-depth study of ensembles is given in Section \ref{subsec:l2r}

Note that we included metrics for listeners who created more than 41+ stations for completeness, even though very few such listeners exists in the dataset. 

Armed with the powerful insight that \RNN~ algorithms perform much better for a subset of listeners, we focus on this set of listeners with > 5 previous stations at prediction time. In Table \ref{tab:rec_filt}, we report the Recall@K and MRR@K results for this filtered set of listeners. \RNN~exhibits large performance gain in both Recall and MRR metrics in all $K$ values over best baseline \KNN. 

\begin{figure}[htb]
\includegraphics[width=.9\linewidth]{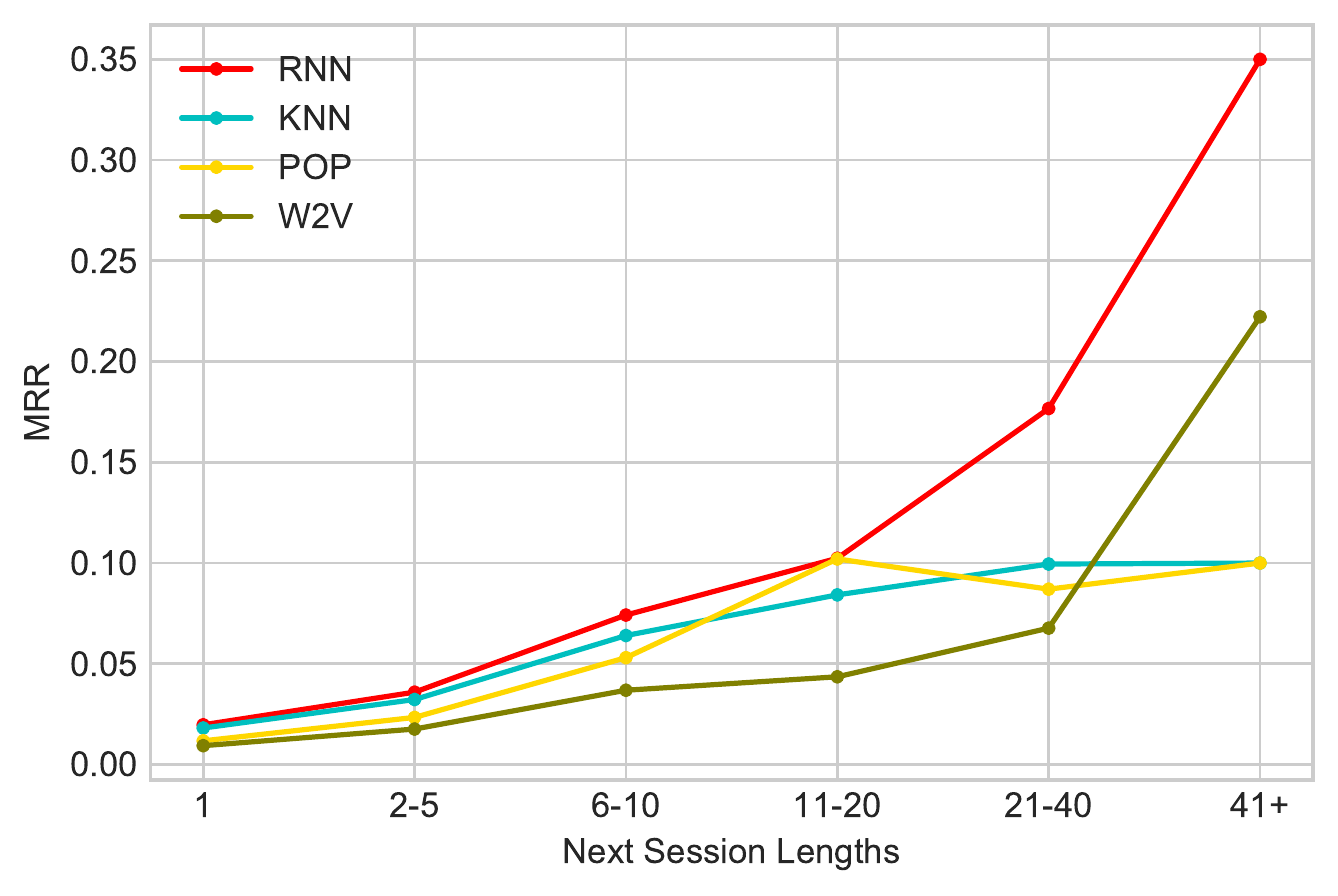}
\caption{MRR@10 \RNN~ and baseline algorithms, segmented by next session length. \RNN~is more effective for listeners who create more stations in the next session.}
\label{fig:mrr_filter_l6_by_next_session}
\end{figure}

We compare performance by segmenting further on the listeners by next session lengths, and show results in Figure \ref{fig:mrr_filter_l6_by_next_session}. The highlight is that the \RNN~ outperforms baselines by a larger and larger margin as the listeners create more stations. This again suggests that \RNN~ learns a sequential dependency that is not captured by other models.

\begin{table}[t]
\centering
\begin{tabular}{l|l|l|l|l|l}
\toprule
\multicolumn{6}{c}{Recall} \\
\midrule
& @$1$ & @$5$ & @$10$ & @$20$ & @$50$ \\
\midrule
\KNN~ & $0.007287$ &  $0.02658$ & $0.04383$ & $0.07049$ & $0.1302$\\
\RNN~ & $0.008121$ &  $0.02897$ & $0.04903$ & $0.08077$ & $0.1436$\\
$\Delta$ \% & $+11.5\%$ & $+8.9\%$ & $+11.9\%$ & $+14.6\%$ & $+10.2\%$ \\
\midrule
\multicolumn{6}{c}{MRR} \\
\midrule
\KNN~ & $0.0134$ & $0.02454$ & $0.02776$ & $0.02978$ & $0.0304$\\
\RNN~ & $0.01485$ & $0.02705$ & $0.03101$ & $0.0335$ & $0.03381$\\
$\Delta$ \% & $+10.9\%$ & $+10.2\%$ & $+11.7\%$ & $+12.5\%$ & $+11.2\%$\\
\midrule
\end{tabular}
\caption{Recall@K and MRR@K for \RNN~ and the best baseline \KNN. Training restricted to listener with more than 5 previous stations.}
\label{tab:rec_filt}
\end{table}

\subsection{Long Tail Music}
\label{subsubsec:long_tail}

In personalized music services, the most difficult task is often to predict long tail (unpopular) music correctly. To further understand the difference among models in this regard, we single out the true positives from the test set for $K=50$, i.e., the intersection of predicted stations and listener created stations, and report results segmented by station popularity. As mentioned in Section \ref{subsec:dataset}, the popularity of a station is represented by the number of occurrences in the training set. Here we bin the occurrence value with a bin size of $500$. 

\begin{figure*}[t!]
\begin{center}
\includegraphics[width=.45\linewidth]{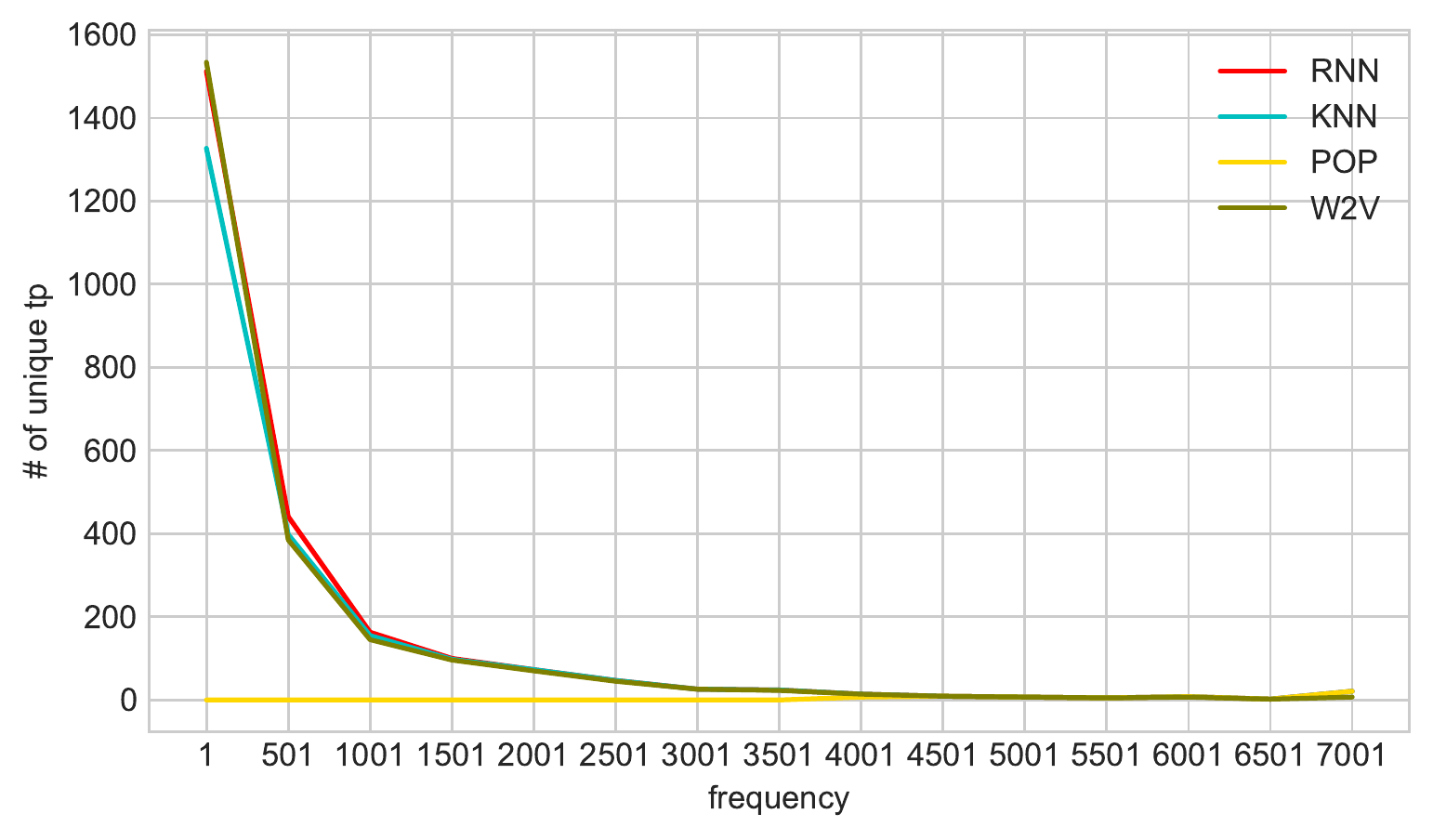}
\includegraphics[width=.45\linewidth]{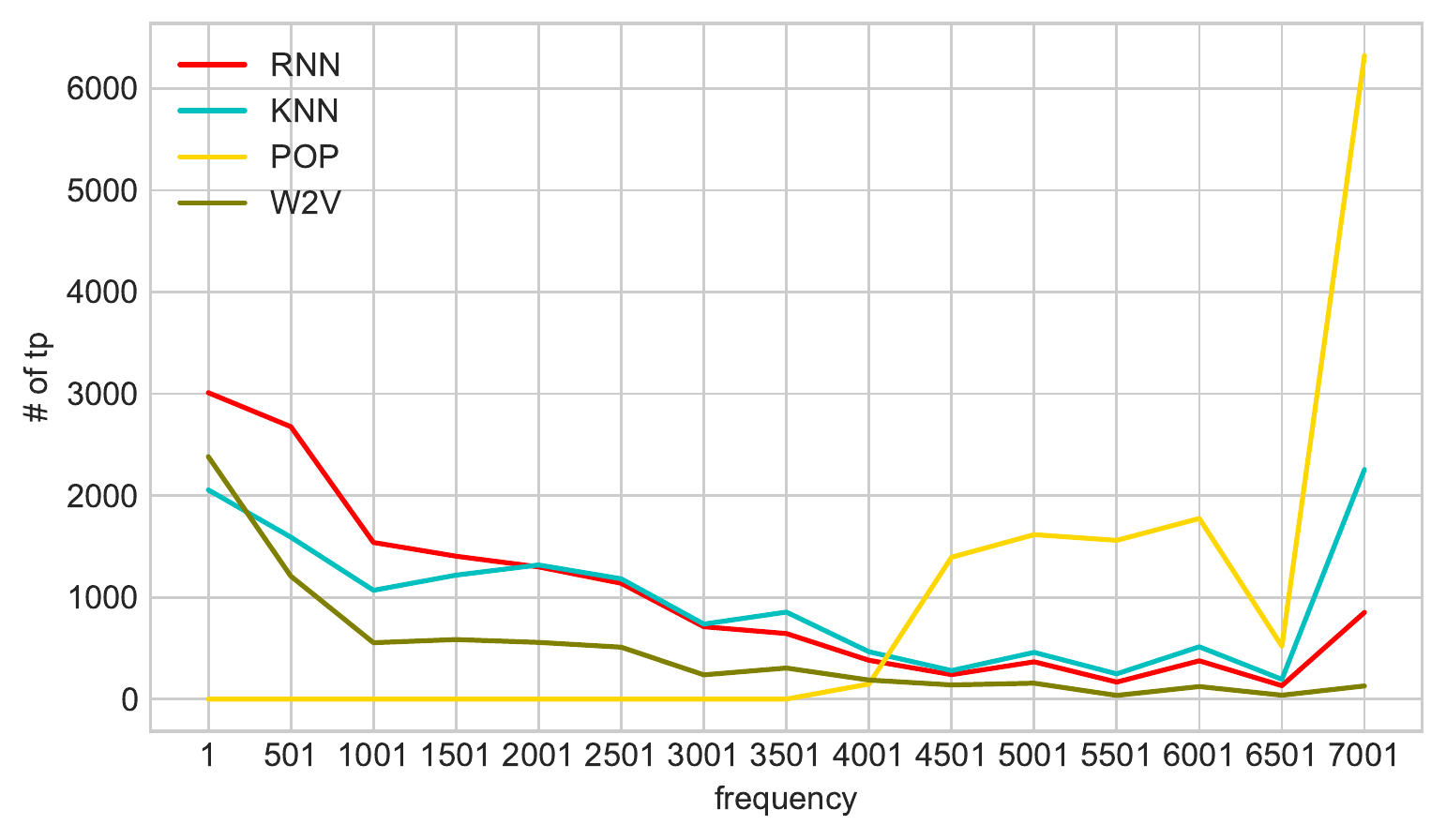}
\caption{Correctly recommended item (true positive) distribution comparison. In the left figure, \RNN~ recommends the most amount of unique least popular items. In the right figure, \RNN~ recommends 50\% more least popular items than \KNN~ in occurrence. }
\label{fig:tp_distr_comp_freq}
\end{center}
\end{figure*}

Not surprisingly, Figure \ref{fig:tp_distr_comp_freq} showed the popularity algorithm predicts more very popular stations correctly. However, this is done at the expense of getting none of the long tail stations right, shown in right of Figure \ref{fig:tp_distr_comp_freq} as the yellow line at $0$ true positives for stations with occurrence < $3500$. Indeed, recall that in the general results in Figure \ref{fig:recall_overall}, \POP~ model lagged behind both the \RNN~ and the \KNN~ models. It is unexpected and very encouraging that \RNN~ in fact recommends almost $50\%$ more least popular stations (those occurred less than $500$ times) accurately than the best baseline, \KNN, as shown in the right of Figure \ref{fig:tp_distr_comp_freq}. In the left figure of Figure \ref{fig:tp_distr_comp_freq}, we also show the \RNN~ and \WTOVEC~ both predict the most unique number of true positives. This provides further evidence that as a listener's taste tends towards more nuanced and less popular, the recurrent neural network captures that tastes evolution more accurately than a simple nearest neighbor approach. Given that the \RNN~ also outperforms all models in Recall@$50$ and MRR@$50$, we believe we demonstrated its value as a promising long tail music recommender.
%
%

\subsection{Learning to Rank}
\label{subsec:l2r}
We lastly look at a combination of our models in a cascaded late fusion ensemble in order to further investigate the predictive power of the \RNN. Recall here that given the stations created by a listener in our training set, our goal is to predict the next stations a listener is going to create. For each listener, we generate the top-$10$ recommendation for each method and consider the union as the set of recommendation candidates to be re-ranked. 

We consider the true target as the station created in the validation set by the same listener. We than label the true targets as relevant and all the others as non-relevant. As the final stage we use  LambdaMart \cite{lambdamart2010} as our supervised ranker to re-rank the candidate list. We tune the LambdaMART model with $500$ trees and the parameters are learned using standard separate training and validation set.

\subsubsection{Features} For this learning to rank (L2R) approach we used a total of $11$ features (a combination of contextual and suggestion features): (1) the inverse of rank of the candidate in \KNN, (2) the inverse of rank of the candidate in \POP~ (3) the inverse of rank of the candidate in \WTOVEC (4) the popularity of the last station created (5) the typology of the last station created (6) the popularity of the candidate station (7) the typology of the candidate station (8) the age of the listener (9) how diverse are the songs listened by the listener (10) how popular are the song listened by the listener (11) the number of stations created by the listener in the training set.

\subsubsection{+RNN} The proposed \RNN~contributes to one additional feature to the previous mentioned one. In this case we include the top-$10$ recommendation generated by \RNN~ to the set of candidates to be re-ranked. We optimize based on MRR@1, MRR@5.

\subsubsection{L2R Results} 
Table \ref{tab:learningtorankimprov} reports the improvement in terms of MRR obtained by using \RNN~ as additional feature in our learning to rank system previously described\footnote{The results on the full dataset show a similar behavior.}. The rows represents two different ways we optimized the system: on metric MRR@1 and metric MRR@5 (note that this is done in the training phase). We then compute the MRR metric on test and report the improvement in percentage terms, particularly MRR@1, MRR@3 and MRR@5. We aim to improve the top ranking results, hence the choice of small $K$ values. Results show an improvement of over $16\%$ in MRR@5 when integrating \RNN~ as additional feature in our learning to rank system. In general, there is a large improvement in all cases presented.

\begin{table}[th]
\centering
\begin{tabular}{ l | l  | l | l } 
 \toprule
 & @$1$ & @$3$ & @$5	$ \\ 
\midrule
  +RNN (opt. MRR@1) & $+10.3\%$ & $+13.8\%$ & $+16.7\%$ \\ 
  +RNN (opt. MRR@5)  & $+9.4\%$ & $+13.5\%$ & $+16.1\%$  \\  
\bottomrule
\end{tabular}
\caption{Learning to rank results. Increase in terms of MRR obtained by using a learning to rank +RNN. We optimize the learning to rank on MRR@1, MRR@5, MRR@10. Results reported are for listeners with historical session length > 5.}
\label{tab:learningtorankimprov}
\end{table}
Together with observations from Section \ref{subsubsec:long_tail}, we interpret this results as a substantial amount of the predictions and the ranking of predictions obtained by \RNN~are complimentary to other models.



\section{Conclusions and future work}
\label{sec:conclusion}
%
This work has presented an investigation into a novel domain in the space of modeling listener music preference dynamics.  We showed the \RNN~model to be extremely effective in modeling the evolution of these preferences, offering a +8.2\% increase in recall@20 over other baselines.

We have found these performance advantages become more pronounced as more historical data is available for a given user.  The RNN outperforms the other baselines most substantially when a listener has between 6 and 40 stations.  This result indicates that long-term dependency structure is important and that the true advantage of the RNN is its ability to encapsulate it.  With less than 6 stations it is perhaps the case that the data is too limited to learn this dependency structure.  The decrease above 40 stations is likely related to a small number of listeners having such high station counts.

One of the most exciting advantages of the RNN is its ability to correctly recommend less popular items.  Items that appeared the least were recommended correctly by the RNN than any other model.

For an additional assessment of the predictive power of the RNN, we also investigated a cascaded learn-to-rank model which allows us to analyze the information gain delivered by the RNN over the other baselines.  We showed a 16\% increase in total performance in MRR@5.

In future work we plan to investigate deeper architectures as well stronger ways to integrate side information.  Architectures that incorporate features like genre and popularity about the station as well as demographics about the listener seem like they could be particularly promising for this work.
Another direction for investigation is to look at the problem in terms of listening behavior instead of station creation.  These models would predict if someone is going to like a particular song. 



\balance
\bibliographystyle{plain}
\bibliography{paper}
\end{document}